\documentstyle[12pt,epsf]{article}
\def\beq{\begin{equation}}
\def\eeq{\end{equation}}
\def\ml{m_\ell}
\def\bq{{\mbox{\boldmath $q$}}}
\def\bp{{\mbox{\boldmath $p$}}}
\def\ap#1#2#3 {Ann. Phys. (NY) {\bf#1} (19#2) #3}
\def\apj#1#2#3 {Astrophys. J. {\bf#1} (19#2) #3}
\def\apjl#1#2#3 {Astrophys. J. Lett. {\bf#1} (19#2) #3}
\def\app#1#2#3 {Acta. Phys. Pol. {\bf#1} (19#2) #3}
\def\ar#1#2#3 {Ann. Rev. Nucl. Part. Sci. {\bf#1} (19#2) #3}
\def\cpc#1#2#3 {Computer Phys. Comm. {\bf#1} (19#2) #3}
\def\err#1#2#3 {{\it Erratum} {\bf#1} (19#2) #3}
\def\ib#1#2#3 {{\it ibid.} {\bf#1} (19#2) #3}
\def\jmp#1#2#3 {J. Math. Phys. {\bf#1} (19#2) #3}
\def\ijmp#1#2#3 {Int. J. Mod. Phys. {\bf#1} (19#2) #3}
\def\jetp#1#2#3 {JETP Lett. {\bf#1} (19#2) #3}
\def\jpg#1#2#3 {J. Phys. G. {\bf#1} (19#2) #3}
\def\mpl#1#2#3 {Mod. Phys. Lett. {\bf#1} (19#2) #3}
\def\nat#1#2#3 {Nature (London) {\bf#1} (19#2) #3}
\def\nc#1#2#3 {Nuovo Cim. {\bf#1} (19#2) #3}
\def\nim#1#2#3 {Nucl. Instr. Meth. {\bf#1} (19#2) #3}
\def\np#1#2#3 {Nucl. Phys. {\bf#1} (19#2) #3}
\def\pcps#1#2#3 {Proc. Cam. Phil. Soc. {\bf#1} (#2) #3}
\def\pl#1#2#3 {Phys. Lett. {\bf#1} (19#2) #3}
\def\prep#1#2#3 {Phys. Rep. {\bf#1} (19#2) #3}
\def\prev#1#2#3 {Phys. Rev. {\bf#1} (19#2) #3}
\def\prl#1#2#3 {Phys. Rev. Lett. {\bf#1} (19#2) #3}
\def\prs#1#2#3 {Proc. Roy. Soc. {\bf#1} (19#2) #3}
\def\ptp#1#2#3 {Prog. Th. Phys. {\bf#1} (19#2) #3}
\def\ps#1#2#3 {Physica Scripta {\bf#1} (19#2) #3}
\def\rmp#1#2#3 {Rev. Mod. Phys. {\bf#1} (19#2) #3}
\def\rpp#1#2#3 {Rep. Prog. Phys. {\bf#1} (19#2) #3}
\def\sjnp#1#2#3 {Sov. J. Nucl. Phys. {\bf#1} (19#2) #3}
\def\spj#1#2#3 {Sov. Phys. JEPT {\bf#1} (19#2) #3}
\def\spu#1#2#3 {Sov. Phys. Usp. {\bf#1} (19#2) #3}
\def\zp#1#2#3 {Zeit. Phys. {\bf#1} (19#2) #3}

\begin{document}
\begin{titlepage}
\begin{center}
{\Large \bf Theoretical Physics Institute \\
University of Minnesota \\}  \end{center}
\vspace{0.2in}
\begin{flushright}
TPI-MINN-98/3-T \\
UMN-TH-1624-98 \\
hep-ph/9803243\\
\end{flushright}
\vspace{0.3in}
\begin{center}
{\Large \bf  Isotopic splittings and OPE in $B \to D$ semileptonic
decays\\}

\vspace{0.2in}

{\bf M.B. Voloshin  \\ }
Theoretical Physics Institute, University of Minnesota, Minneapolis,
MN
55455 \\ and \\
Institute of Theoretical and Experimental Physics, Moscow, 117259
\\[0.2in]
\end{center}

\begin{abstract}

It is shown that the kinematical difference in the decays $B^- \to D^0
\, \ell \, \tilde \nu$ and $B^0 \to D^+ \, \ell \, \tilde \nu$ due to
the isotopic mass splittings of the $B$ and $D$ mesons is compensated in
the total decay rate by an appropriate difference in the lepton spectra.
Thus there is no effect on the total decay rates in the linear order in
the isotopic mass splittings, as required by the general consideration
based on the operator product expansion. Although phenomenologically the
isotopic difference in the spectra amounts to at most about 1\%, in the
theoretical aspect this effect can be viewed as an additional
illustration of how the general OPE results emerge from the properties
of exclusive channels.
\end{abstract}
\end{titlepage}

It has become a matter of common knowledge\footnote{For a recent review
see e.g. Ref. \cite{review}.} that the inclusive rates of weak decays
for hadrons containing a heavy quark are governed by the short-distance
QCD$^{\cite{sv0}}$ and are given by the decay rate of the heavy quark.
The corrections to this leading behavior are suppressed by at least two
powers of the inverse heavy quark mass, $m_Q^{-2}$, or, in the situation
where two heavy quarks are involved, like in the $b \to c$ transitions,
by the inverse second power of some combination of $m_b$ and $m_c$:
$m_{b,c}^{-2}$. Moreover,  the dependence on the flavors of the light
quarks in the hadron arises only starting with the subsequent
order$^{\cite{sv0,bgt,vs}}$ $m_{b,c}^{-3}$. On the other hand in the
so-called SV limit$^{\cite{sv}}$: $m_{b, c} \to \infty$,
$\Lambda_{QCD} \ll \Delta \equiv m_b-m_c \ll m_{b,c}$, the inclusive
rate of the semileptonic decays $B \to X_c \, \ell \, \tilde \nu$ is
saturated by just one exclusive decay channel: $B \to D \, \ell \,
\tilde \nu$ for the vector part of the $b \to c $ current and $B \to D^*
\, \ell \, \tilde \nu$ for the axial part of the current.
The latter property comes into effect due to that the form factor $F$ of
e.g. the vector $b \to c$ current: $\langle D | c^\dagger \, b| B
\rangle$ is equal to one$^{\cite{sv}}$, $F=1$, in the limit $m_{b, c}
\to \infty$ at zero recoil of the $D$ meson, and the zero recoil limit
is the only one relevant for calculating the total rate of the $B \to D
\, \ell \, \tilde \nu$ decay if $\Delta \ll m_{c}$. The relation $F=1$
is subject to small and calculable QCD radiative
corrections$^{\cite{sv,cj}}$ that match the same corrections to the
parton decay rate of the quark: $b \to c \, \ell \, \tilde \nu$, and at
large but finite masses of the heavy quarks the mass corrections are
also suppressed by at least$^{\cite{sv,luke}}$ $m_{b,c}^{-2}$. The total
rate of the decay is then given by the well known expression:
\beq
\Gamma (B \to D \, \ell \, \tilde \nu) = \eta \, {G_F^2 \, |V_{cb}|^2 \,
\Delta^5 \over 60 \, \pi^3}~~,
\label{g0}
\eeq
where $\eta$ is the QCD radiative correction factor, and, for
simplicity, the mass of the lepton is assumed to be small, $\ml \to 0$.
According to the general consideration of the heavy quark theory, the
masses of the $B$ and $D$ mesons are heavier than the corresponding
quark masses by equal amount (up to terms of order $m_Q^{-2}$):
$M_{B,D}= m_{b,c} + \overline \Lambda + {\cal O} (m_{b,c}^{-2})$, so
that the energy release $\Delta$ in eq.(\ref{g0}) is the same in the
meson and in the quark decay: $M_B-M_D = m_b-m_c + {\cal O}
(m_{b,c}^{-2})$.

This agreement between the exclusive decay rate and the inclusive one
comes into question if one takes into account the isotopic mass
splittings of the $B$ and $D$ mesons. Indeed, the energy release in the
decay $B^- \to D^0 \, \ell \, \tilde \nu$ is different from that in $B^0
\to D^+ \, \ell \, \tilde \nu$ by a small but non-zero amount
\beq
\delta m= \left [ M(B^-) -M(D^0) \right ] - \left [ M(B^0) -M(D^+)
\right ] = \left [ M(D^+) -M(D^0) \right ] - \left [ M(B^0) -M(B^-)
\right ].
\label{d2m}
\eeq
Thus naively applying the formula in eq.(\ref{g0}) to these exclusive
decays would give a relative difference in the rates of these decays
$\delta \Gamma / \Gamma= 5 \, \delta m /\Delta$ scaling as inverse {\em
first} power of $\Delta$. If identified with the total semileptonic
decay rate in the SV limit, this scaling behavior of the isotopic
correction would contradict to the general short-distance OPE
description of the inclusive decay, since within the OPE there are no
operators of the appropriate dimension. Moreover, the dependence of the
rates on the spectator quark flavor may arise in OPE only starting from
the terms scaling as inverse third power of the heavy quark masses. It
is the purpose of this paper to show how the isotopic difference in the
energy release is reconciled with the general results from the OPE. The
final answer turns out to be that due to the electromagnetic (in fact
Coulomb) interaction of the charged lepton $\ell$ with the spectator
quark there also arises an isotopic difference in the lepton spectrum of
the exclusive decays. At energy of the lepton $E$, such that $E \gg
\Lambda_{QCD}$, this difference is related to $\delta m$ as
\beq
{\delta \, d\Gamma/dE \over d\Gamma/dE} = -\delta m \, {2 E^2 - \ml^2
\over E \, (E^2 - \ml^2)}~~.
\label{res}
\eeq
After integration over the energy this spectral correction completely
cancels in the total rate the correction due to the isotopic difference
in the energy endpoint in the order $\delta m /\Delta$. The
cancellation, as expected, does not depend on either the charged lepton
mass, or other ingredients, e.g. the neutrino mass, that one might
choose to introduce for the purpose of a theoretical cross-check.

For the actual $B$ and $D$ mesons the discussed mass splitting
is$^{\cite{pdg}}$ $4.43 \pm 0.31$ MeV, thus at the typical energy of the
charged lepton $E \approx 1$ GeV, the isotopic difference in the lepton
spectra amounts to about 0.9\% (at $\ml \to 0$) and is smaller at higher
energy. Towards lower energies the condition $E \gg \Lambda_{QCD}$,
necessary for deriving  eq.(\ref{res}), starts to be invalidated, and
the spectral difference generally is not expressed in terms of $\delta
m$, but rather becomes sensitive to details of the electromagnetic form
factor of the heavy mesons, and therefore may be used for a study of
this form factor, provided that sufficiently precise experimental data
may become available. In this paper however we are primarily concerned
with the theoretical aspect of accommodating the isotopic differences
between exclusive channels within the general OPE approach.

Proceeding to details of the argument,  we first take a closer look into
the isotopic mass splittings within the heavy quark theory. The leading
dependence of the meson mass on the flavor of the light quark $q$
appears within the heavy mass expansion in the flavor dependence of the
parameter $\overline \Lambda$:
\beq
M(Q \bar q)=m_Q + {\overline \Lambda}_q + {\cal O} (m_Q^{-2})~,
\label{genl}
\eeq
where the effects of the light quark mass and of the electromagnetic
interaction within the meson can be parametrized in ${\overline
\Lambda}_q$ as
\beq
{\overline \Lambda}_q={\overline \Lambda}_0 + \mu_q -\alpha \, Q_Q \,
Q_q \mu_C + \ldots
\label{lamq}
\eeq
In this equation ${\overline \Lambda}_0$ stands for the value of
${\overline \Lambda}$ in the limit of massless spectator quark and of
zero electric charges of the quarks, $\mu_q$ is the shift of the meson
mass due to the light quark mass:
\beq
\mu_q=m_q \, {\partial M(Q \bar q) \over \partial m_q} = \langle (Q \bar
q) | m_q \, (\overline q \, q) | (Q \bar q) \rangle~~,
\label{muq}
\eeq
and the term with $\mu_C$ is due to the electromagnetic (essentially
Coulomb) interaction between the quarks in the meson, where $\alpha$ is
the QED fine structure constant, and $Q_Q$ and $Q_q$ are the electric
charges of the quarks in units of $|e|$. Finally, the ellipses in
eq.(\ref{lamq}) stand for higher order terms in $\alpha$ and $m_q$ and
those terms will be completely ignored in what follows. It is clear from
eq.(\ref{lamq}) that in the isotopic difference of the energy release
in the $B^- \to D^0$ and $B^0 \to D^+$ transitions, $\delta m$ (cf.
eq.(\ref{d2m})), the terms with $\mu_u$ and $\mu_d$ cancel, and the
resulting effect is purely electromagnetic:
\beq
\delta m = \alpha \, (Q_c-Q_b) \, (Q_u-Q_d) \, \mu_C~~.
\label{d2me}
\eeq

In the limit of heavy mass $m_Q$ the quantity $\mu_C$ can be expressed
in terms of the form factor $f_q$ of the vector current of the light
quark $j_\mu^{(q)}=(\bar q \, \gamma_\mu \, q)$ for the meson:
\beq
\langle (Q \bar q) |\, j_\mu^{(q)} (\bq) \, | (Q \bar q) \rangle =
-f_q(\bq^2) \, \delta_{\mu 0}~~,
\label{fq}
\eeq
where the nonrelativistic normalization for the heavy states is used,
and the condition $|\bq| \ll m_Q$ is implied, so that the spatial
components of the current are small: $-f_q(\bq^2) \, \bq /m_Q$, and the
recoil energy effects ${\cal O}(\bq^2/m_Q^2)$ can also be ignored.
The shift of the energy proportional to $Q_Q Q_q$ is then identified as
the cross term between the electromagnetic currents of the heavy $Q$
quark and the light $q$ in the general formula for the electrostatic
energy, so that one finally finds
\beq
\mu_C= 4\pi \, \int \, {f_q (\bq^2) \over \bq^2} \, {d^3 q \over (2
\pi)^3}~~.
\label{mufq}
\eeq
In this equation it is taken into account, that at $|\bq| \ll m_Q$ the
heavy quark has only electrostatic interaction in the leading order in
$m_Q^{-1}$, and that the form factor of the heavy quark current is equal
to one in this approximation. It should be also noticed that $f_q$ is
defined here as the form factor for the spectator quark flavor and is
normalized as $f_q(0)=1$, thus the difference between the heavy mesons
with different flavor of the light antiquark arises only through the
overall factor of the spectator electric charge. (Taking into account
flavor differences in the $f_q$ itself would lead to terms of higher
order in the isospin, or the flavor SU(3), breaking.)

The assumption that it is sufficient to consider only the region of
momenta such that $|\bq| \ll m_Q$, where the heavy quark is static, is
consistent as long as the integral in eq.(\ref{mufq}) is convergent.
This definitely is the case, since the asymptotic behavior of the form
factor $f_q$ at $\bq^2 \gg \Lambda_{QCD}^2$ (but still $\bq^2 \ll
m_Q^2$) can be deduced from the short-distance QCD: $f_q(\bq^2) \sim
\alpha_s(\bq^2)/|\bq|^3$.

It can be also noticed that the derivation of eq.(\ref{mufq}) does not
rely on the assumption (generally incorrect) that the overall
electromagnetic form factor of the $Q \bar q$ meson is simply a sum of
the electromagnetic form factors of the $Q$ quark and the $\bar q$
antiquark. It is only the light-flavor-dependent part of the
electromagnetic energy of the meson, which is related by the equation
(\ref{mufq}) to the cross term between these form factors.

Returning to the semileptonic $B \to D$ decays, we discuss the new
ingredients brought in by inclusion of the electromagnetic interaction,
which breaks the heavy quark symmetry$^{\cite{iw}}$ and results in a
number of corrections. On the OPE side, i.e. in terms of the parton
decay of the $b$ quark, there arise QED radiative corrections. These
corrections are known$^{\cite{am}}$ and are determined only by $Q_b$ and
$Q_c$, thus being insensitive to the isotopic charge splitting of the
light quarks $Q_u - Q_d$. Moreover, these corrections match those for
the exclusive $B \to D \, \ell \, \bar \nu$ decays and are not essential
for the present discussion. Thus within the OPE approach it is required
that the linear in the isotopic splitting effects should vanish in the
total semileptonic decay rates of the $B$ mesons.

On the exclusive side, i.e. considering the modifications of the decay
into exclusive channels, the difference in the electric charges
generally violates the heavy quark symmetry relation $F=1$ for the form
factor of the weak $B \to D$ transitions in the SV limit. However, the
effect on the $F$ due to the electrostatic mass shift is readily
verified to be of the second order in the electrostatic energy, in
agreement with the general Ademollo-Gatto theorem. Additionally, there
arise in the first order in the electrostatic energy the non-vanishing
amplitudes of transitions of the $B$ mesons into inelastic states, i.e.
into the states different from the ground-state $D$ mesons. However, the
effect of these transitions in the total rate is obviously of the second
order in the isotopic splittings.

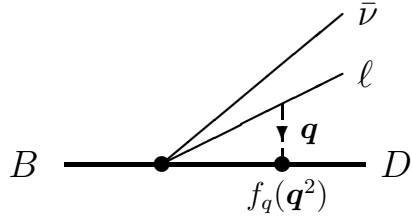
\begin{figure}
\thicklines
\unitlength 1mm
\begin{picture}(61.00,30.00)(-30.00,5.00)
\put(20.00,16.00){\line(1,0){40.00}}
\put(20.00,15.80){\line(1,0){40.00}}
\put(33.00,16.00){\line(6,5){24.00}}
\put(33.00,16.00){\line(2,1){24.00}}
\put(49.00,16.00){\line(0,1){2.00}}
\put(49.00,21.00){\vector(0,-1){2.00}}
\put(49.00,22.00){\line(0,1){2.00}}
\put(18.00,16.00){\makebox(0,0)[rc]{{\large $B$} }}
\put(62.00,16.00){\makebox(0,0)[lc]{{\large $D$}}}
\put(50.00,20.00){\makebox(0,0)[lc]{ \bq}}
\put(59.00,28.00){\makebox(0,0)[lc]{{\large $\ell$}}}
\put(59.00,36.00){\makebox(0,0)[lc]{{\large $\bar \nu$}}}
\put(33.00,16.00){\circle*{2.00}}
\put(49.00,16.00){\circle*{2.00}}
\put(49.00,14.00){\makebox(0,0)[tc]{ $f_q(\bq^2)$}}
\end{picture}
\caption{The graph for the light-flavor-dependent part or the QED
correction to the decay $B \to D \, \ell \, \bar \nu$. The dashed line
denotes the photon, which in the SV limit is purely a Coulomb one carrying the spatial momentum \bq.}
\end{figure}

In order to find the sought linear in the isotopic splitting effect in
the exclusive decay one should look into the light flavor dependent part
of the QED radiative corrections in the decays $B \to D \, \ell \, \bar
\nu$. In the SV limit this part arises through the difference of the
Coulomb interaction of the charged lepton $\ell$ with the produced $D$
meson, shown in Figure 1. Clearly, this difference is proportional to
$Q_\ell \, (Q_u-Q_d)$, and the charge of the lepton is related to the
charges of the heavy quarks by the charge conservation:
$Q_\ell=Q_b-Q_c$, thus giving the correct parametric dependence of the
effect on the quark electric charges. Furthermore, in the SV limit, and,
as will be seen, given the convergence of the integral in
eq.(\ref{mufq}), it is sufficient to consider only static $B$ and $D$
mesons, which produce only the Coulomb field. The difference
\beq
\delta \, {d\Gamma \over dE}={d \Gamma(B^- \to D^0 \, \ell \, \bar \nu)
\over dE}- {d \Gamma(B^0 \to D^+ \, \ell \, \bar \nu) \over dE}
\label{deldef}
\eeq
of the spectra in the energy $E$ of the charged lepton is generated by
the interference of the graph of Fig.1 with the `bare' amplitude and is
readily found in the form:
\beq
\delta \, {d\Gamma \over dE} = -  {d\Gamma \over dE} \, 4 \pi \, \alpha
\, (Q_c-Q_b) \, (Q_u - Q_d)\, { 1 \over E} \, 2 \, {\rm Re} \left [ \int
\, {f_q(\bq^2) \over \bq^2} \, {2 \, E^2 + (\bp \cdot \bq) \over \bq^2+2
\, (\bp \cdot \bq) } \, {d^3 q \over (2 \pi^3)} \right ]~~,
\label{delf1}
\eeq
where $\bp$ is the momentum of the charged lepton and no assumption is
made about the mass of the neutrino or of the charged lepton, so that in
particular $E=\sqrt{\bp^2 + \ml^2}$.

In calculating the integral in eq.(\ref{delf1}) one can replace the
factor with the angular dependence  by its average over the angle
between $\bp$ and $\bq$:
\beq
\left \langle {2 \, E^2 + (\bp \cdot \bq) \over \bq^2+2 \, (\bp \cdot
\bq) } \right \rangle =
{1 \over 2} + { E^2 - q^2/4  \over 2 \, p \, q } \ln {q+2p \over
q-2p}~~,
\label{av}
\eeq
where $p=|\bp|$ and $q=|\bq|$. Let us now impose the condition that $p
\gg \Lambda_{QCD}$ and thus $p$ is much larger than the characteristic
values of $q$ in the form factor $f_q (q^2)$. This allows to consider
the expansion of the expression in eq.(\ref{av}) at small $q$. The
leading term in the expansion in $q$ is proportional to $q^{-1}$ and is
purely imaginary\footnote{This purely imaginary term develops into the
difference of the Coulomb scattering phases, that is logarithmically
divergent in the infrared and is proportional to $f_q(0)=1$.}, while the
leading contribution to the real part is independent of $q$ and is given
by:
\beq
2 \, {\large \rm Re} \left \langle {2 \, E^2 + (\bp \cdot \bq) \over
\bq^2+2 \, (\bp \cdot \bq) } \right \rangle={2 E^2 - \ml^2 \over E^2 -
\ml^2} + {\cal O}\left ( {q^2 \over p^2} \right )~~.
\label{reav}
\eeq
The higher terms in this expansion result in corrections to the total
rate scaling as higher powers of $\Delta^{-1}$ that have corresponding
terms in the OPE. Thus of interest in the present discussion is only the
leading term in eq.(\ref{reav}). Keeping only this leading term, one
readily rewrites the spectral difference of eq.(\ref{delf1}) in the form
\beq
\delta \, {d\Gamma \over dE} = -  {d\Gamma \over dE} \, 4 \pi \, \alpha
\, (Q_c-Q_b) \, (Q_u - Q_d)\,{2 E^2 - \ml^2 \over E \,  (E^2 - \ml^2)}
\int \, {f_q(\bq^2) \over \bq^2} \, {d^3 q \over (2 \pi^3)}~~.
\label{delf2}
\eeq
By comparing this expression with the equations (\ref{d2me}) and
(\ref{mufq}) one arrives at the final result in eq.(\ref{res}) for the
isotopic difference of the charged lepton energy spectra.

It can now be shown explicitly that the difference in the spectra
exactly cancels the effect of the isotopic difference in the total
energy release $\Delta$. Indeed, the total rate calculated as an
integral over the `bare' energy spectrum can be written as
\beq
\Gamma_0= \eta \, {G_F^2 \, | V_{cb} |^2 \over 2 \, \pi^3}
\int_{\ml}^{E_{max}} \, E_\nu \, p_\nu \, E \, p \, dE~~,
\label{gs0}
\eeq
where $E_\nu=\Delta-E$  and $p_\nu=\sqrt{E_\nu^2-m_\nu^2}$ are the
energy and the momentum of the neutrino, and we allow for an arbitrary
non-zero neutrino mass in order to illustrate the robustness of the
discussed cancellation. The upper limit of integration is then
$E_{max}=\Delta-m_\nu$. Notice also that the QCD correction factor  in
the SV limit is constant over the spectrum$^{\cite{sv}}$. Let us
introduce the notation $N(E_\nu)=E_\nu \, p_\nu$ and note that it enters
in eq.(\ref{gs0}) as $N(\Delta-E)$ and that $N(\Delta-E_{max})=0$. Due
to the latter property the linear in $\delta m$ change in the rate under
the shift of $\Delta$: $\Delta \to \Delta + \delta m$ is given by:
\beq
\delta_1 \Gamma= \eta \, {G_F^2 \, | V_{cb} |^2 \over 2 \, \pi^3} \,
\delta m \, \int_{\ml}^{E_{max}} \left [ {d N(\Delta-E) \over d\Delta}
\right ] \, E \, p \, dE ~~.
\label{delta1}
\eeq
On the other hand, using the elementary relation
$$
E \, p(E) \, {2 E^2 - \ml^2 \over E \,  (E^2 - \ml^2)} = {d \over dE}\,
E \, p(E)~~,
$$
we find from eq.(\ref{res}) for the change of the rate due to the
isotopic difference in the lepton spectra the expression
\beq
\delta_2 \Gamma= - \eta \, {G_F^2 \, | V_{cb} |^2 \over 2 \, \pi^3} \,
\delta m \, \int_{\ml}^{E_{max}}\, N(\Delta - E) \, \left [ {d \over
dE}\, E \, p(E) \right ] \, dE = -\delta_1 \Gamma~~,
\label{delta2}
\eeq
where the latter transition involves integration by parts (with the
relations $N(\Delta-E_{max})=0$ and $p|_{E=\ml}=0$ taken into account)
and also noticing that $d N(\Delta -E)/dE = - dN(\Delta-E)/d\Delta$.
Thus we find that $(\delta_1+\delta_2) \Gamma =0$, which concludes our
proof that the total semileptonic rate is not affected by the isotopic
mass differences in the linear order in $\delta m$, independently of the
charged lepton or the neutrino masses, in full compliance with the OPE
result. It should be also mentioned that although for notational
definiteness the reasoning above is given for the decays of the $B$
mesons into the pseudoscalar $D$ mesons, all the formulas are fully
applicable to the $B \to D^*$ transitions, since in the SV limit the
discussed effects are determined by the spin-independent electrostatic
interaction.

The isotopic difference in eq.(\ref{res}) is found in this paper in the
SV limit, where the heavy mesons in the $B \to D \, (D^*) \, \ell \,
\bar \nu$ decays are strictly static. The theoretical parameter
$\xi=(m_b-m_c)^2/(m_b+m_c)^2$ governing the deviation from this limit
for the actual $B \to D \, (D^*)$ transitions$^{\cite{sv}}$ is not very
small: $\xi \approx 0.3$. Also phenomenologically it is known that
the exclusive decays $B \to D \, (D^*) \, \ell \, \bar \nu$ saturate
about 65\% of the total semileptonic rate, rather than completely, as
they should in the SV limit. Therefore a more elaborate consideration
beyond the static SV approximation is desirable for more accurate
predictions of the experimentally measurable isotopic difference in the
lepton spectra, if such experimental study appears on the agenda. At
this point, based on eq.(\ref{res}) and on the reasonable smallness of
$\xi$, one can assert that the difference in the spectra is not
hopelessly small and should amount to a sizeable fraction of 1\% at the
lepton energy about 1 GeV. At smaller energies, where the condition $p
\gg \Lambda_{QCD}$ cannot be used, the full formulas in
eqs.(\ref{delf1}) and (\ref{av}) and their modification beyond the SV
can be used for a study of the form factor $f_q(q^2)$ in the spacelike
region.

I am thankful to Arkady Vainshtein and Yuichi Kubota for enlightening
discussions of the theoretical aspects and experimental possibilities.
This work is supported in part by the DOE  grant  DE-FG02-94ER40823.

\end{document}